\documentclass[prl,showpacs,floatfix,twocolumn]{revtex4}
\usepackage{epsfig}

\begin{document} 
\title{Coulomb blockade of a noisy metallic box: A realization of Bose-Fermi
Kondo models}

\author{Karyn Le Hur}
\affiliation{$^1$ D\'epartement de Physique and RQMP,
 Universit\'e de Sherbrooke, Sherbrooke, Qu\'ebec, Canada, J1K 2R1}

\newcommand{\br}{{\bf r}}
\newcommand{\ovl}{\overline}
\newcommand{\hw}{\hbar\omega}
\newcommand{\mybeginwide}{
    \end{multicols}\widetext
    \vspace*{-0.2truein}\noindent
    \hrulefill\hspace*{3.6truein}
}
\newcommand{\myendwide}{
    \hspace*{3.6truein}\noindent\hrulefill 
    \begin{multicols}{2}\narrowtext\noindent
}
 
\begin{abstract}
We focus on a metallic quantum dot 
coupled to a reservoir of electrons through a single-mode point contact 
and capacitively connected
to a back gate, by including that the gate voltage 
can exhibit {\it noise}; this will occur when connecting the gate
lead to a transmission line with a finite 
impedance. The voltage  
fluctuations at the back gate can be described
through a Caldeira-Leggett model of harmonic oscillators. For {\it weak} 
tunneling between the lead and the dot, exploiting the anisotropic 
Bose-Fermi spin model, we show that zero-point fluctuations of the 
environment can markedly alter the Matveev Kondo fixed
point leading to an amplification of the charge quantization phenomenon.
\end{abstract}

\maketitle

The single-electron box, the simplest single-electron circuit, is probably
the ideal system in which to test the theory of quantum charge fluctuations.
The box has been vividly investigated theoretically because
this is a model system for understanding electron-electron interactions and 
because the quantum fluctuations in the box are analogous to the Kondo 
effect\cite{Matveev}. In practice, charge fluctuation measurements 
necessitate a (large) 
dot at the micron scale 
with a very dense spectrum\cite{Ashoori}; This is referred 
to as a {\it metallic dot}.
In earlier theoretical
treatments of the metallic dot coupled to a 
reservoir lead\cite{Matveev,Grabert}, the 
gate voltage was 
treated as a fixed parameter of the Hamiltonian. As depicted in Fig. 1, 
in this Letter we consider the 
more general situation where the source of the gate voltage
is placed in series with an impedance Z$(\omega)$ of the gate 
lead showing resistive behavior at low frequency. From
the fluctuation-dissipation theorem this resistance will introduce
noise in the gate voltage even at zero temperature. Below, dot and lead are 
coupled through a single-mode quantum point contact (QPC).

\vskip -0.05cm
Since the charge 
on a quantum box can now be measured up to few 
thousandths of an electron\cite{Konrad}, it is then important to ask
to what extent zero-point fluctuations of the electrical environment 
affects the (Kondo) ground state energy, i.e., the capacitance, 
of a metallic dot already coupled to a reservoir of electrons through a  
single-mode
QPC\cite{Matveev}. While the transport 
through a single ultrasmall tunnnel junction coupled to an electromagnetic 
(dissipative) environment
has been thoroughly investigated\cite{Nazarov2}, to our knowledge 
a discussion for the 
Coulomb blockade oscillations of a metallic box subject to a fluctuating gate 
voltage 
seems not to be available. According to conventional wisdom, 
{\it tunneling rates of quasiparticles become
strongly suppressed in the presence of dissipative 
environments because an additional energy is needed to excite the
environment}\cite{Devoret}; Hence, 
this should certainly modify the 
Kondo picture ascribed by Matveev for electron tunneling between the 
island and the lead\cite{Matveev}. 

\vskip -0.05cm
The back gate here is embodied
by a voltage fluctuating around its average value 
$V_g$. We treat the full dynamics of the fluctuations
in a self-consistent manner; $\delta V_g(t)$ representing 
the voltage fluctuations
(or the charge dis-

\begin{figure}[ht]
\centerline{\epsfig{file=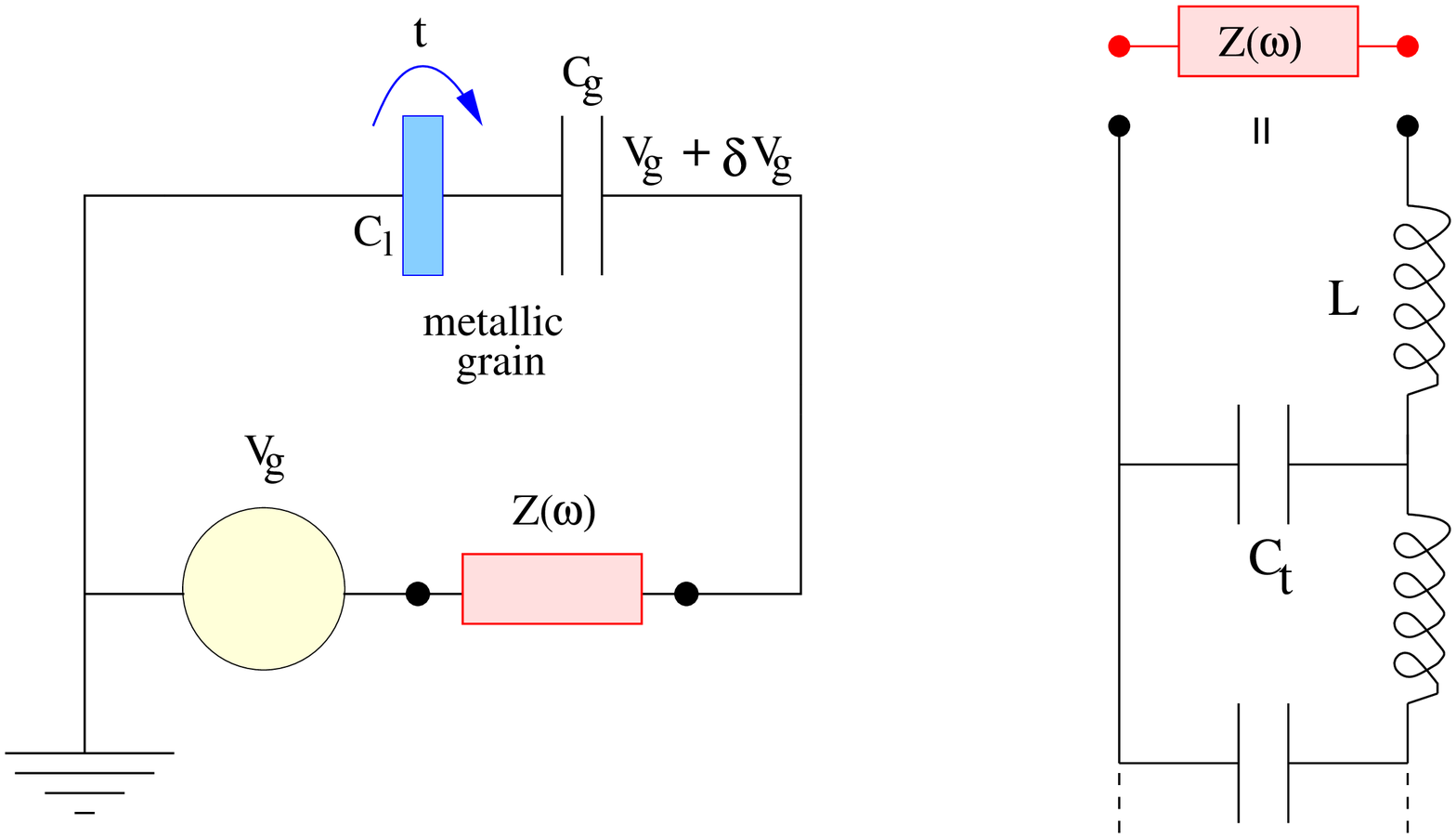,angle=0.0,height=4.9cm,width=
8.3cm}}
\vskip -0.2cm
\caption{Circuit diagram of the metallic dot coupled to a bulk lead and 
subject to a fluctuating back-gate voltage $V_g+\delta V_g(t)$. The 
circuit is supposed to exhibit 
an Ohmic resistance $R=Z(\omega=0)$. We incorporate the finite
resistance in a Hamiltonian fashion, 
through a long dissipative transmission line with capacitance
$C_t$ and inductance $L$: $R=(L/C_t)^{1/2}$.}
\vskip -0.5cm
\end{figure}

\hskip -0.3cm -placement up to a factor) on the back gate
is assumed to be given by the sum over 
the normal coordinates of a Caldeira-Leggett type bath of harmonic 
oscillators\cite{CL}:
\begin{equation}\label{bath}
H_B=\sum_{j=1}^{+\infty} \left(\frac{{{P}_j}^2}
{2M}+\frac{M{\omega_{j}}^2}{2}{X_j}^2\right),
\end{equation}
and
\begin{equation}
\delta V_g(t)=\gamma\sum_{j}\lambda_j X_j(t).
\end{equation}
This mapping is entirely justified in the context of a long dissipative
transmission line; The precise correspondance\cite{note0} between mechanical 
and electrical quantities can be found, e.g., in the recent 
Ref.~\onlinecite{Markus}. In the 
case of a linear circuit element with Ohmic resistance $R$, one 
expects Johnson-Nyquist correlation functions 
$\langle \delta V_g  \delta V_g\rangle_{\omega}\approx \hbar R\omega \coth\left(\hbar
\omega/2k_B T\right)$; This usually refers to as an Ohmic dissipative 
environment. For convenience, we have extracted the {\it dimensionless} 
parameter $\gamma=\sqrt{R/R_K}$ which naturally 
characterizes the coupling between the bosonic 
bath and electrons of the box; $R_K=h/e^2\approx 25.8k\Omega$
being the quantum of resistance. Having in mind a long dissipative transmission
line with capacitance $C_t$ and inductance $L$, the Ohmic resistance is
$R=\sqrt{L/C_t}$. Furthermore, the couplings $\lambda_j$, the mass 
and the frequencies of the
oscillators enter only through the bath's spectral density
$J(\omega)=\gamma^2\sum_j {\lambda_j}^2 
\pi\delta(\omega-\omega_j)/(2M\omega_j)=\hbar R\omega$ in the quantum limit; $\omega_c=\sqrt{1/(LC_t)}$ is a natural low-frequency cutoff 
for the transmission line. 
Using the circuit of Fig.~1, the Coulomb term for the quantum box 
takes the form:
\begin{equation}
H_c=E_c(Q-N)^2-E_cN^2-eQ\delta V_g.
 \end{equation}
Here, $eN=V_g C_g$ is proportional to the mean gate voltage and
$eQ$ depicts the charge on the dot. Here, the charging energy
of the granule takes the form $E_c=e^2/(2 C_{\Sigma})$ where
$C_{\Sigma}=C_g+C_l$; $C_g$ denotes the 
classical capacitance between the back-gate and the grain and $C_l$ is
the capacitance between the lead and the island.

Let us first study the case of a low-transparency tunnel 
barrier and focus on the point N=1/2 where the states with
Q=0 and Q=1 are energy degenerate, and charge fluctuations in
principle can be large. We introduce the small parameter $h=eV_g-e^2/(2 C_g)
\propto (N-1/2)$ to measure deviations of N from the degeneracy point. Our 
treatment of electron tunneling is based on the assumption
that the tunneling Hamiltonian takes the form\cite{Matveev}
\begin{equation}
H_t=\sum_{k,q} |t| c^{\dagger}_{dk} c_{lq} S^+ +h.c.,
\end{equation}
where $|t| c^{\dagger}_{dk}c_{lq}$ is the tunneling term transferring
an electron from the lead (l) to the (large) dot (d), and where $S^+$
is an operator raising the charge $Q$ on the dot from 0 to 1. 
The tunneling matrix element $t$ is assumed not to depend on the momenta
$k$ and $q$. Below, we will discuss in detail the case of {\it spinless} 
electrons, assuming that a strong in-plane magnetic field has been applied, 
and as mentioned previously the case of a point contact
most probably with 
{\it one} conducting transverse mode\cite{Matveev,Georg}. 
We are 
led to the obvious identification $Q=S^z+1/2$, and finally to
\begin{equation}
\label{ising}
H_c=-hS^z-\gamma S^z\Phi;
\end{equation}
$\Phi(t)=e\sum_j \lambda_j X_j(t)$ now stands for the bosonic variable.
It is also appropriate to visualize $H_t$ as a transverse Kondo Hamiltonian
$H_t=(J_{\perp}/2)\left(s^+S^-+h.c.\right)$ where $J_{\perp}=2|t|$ and $s^+=
\sum_{k,q} c^{\dagger}_{lq}c_{dk}$. Bear in mind that here a Kondo 
flip process means physically 
the transfer of an electron from lead
to dot or vice-versa, the effective spin index $\alpha=l,d$ 
is in fact the position of an electron in the structure, and then 
$s^z=\sum_{k,q}
\left(c^{\dagger}_{lk}c_{lq}-c^{\dagger}_{dk}c_{dq}\right)$. 

Hence, in the limit
of weak tunneling $(|t|\rho)\ll 1$, our system is embodied by the
Bose-Fermi spin Hamiltonian
\begin{eqnarray}
\label{hami}
H &=& \left(\sum_{\alpha=l,d} H_{Kin}^{\alpha}-h S^z\right) +H_B
\\ \nonumber
 &+& \frac{J_{\perp}}{2}\left(s^+S^-+h.c.\right)-\gamma\Phi S^z;\\ \nonumber
\end{eqnarray}
$H_{Kin}^{\alpha}$ stand for the usual 
kinetic terms in the lead and dot.
The densities of states in the dot and lead have been
assumed to be equal for simplicity, and are denoted $\rho$. It is 
suitable
to recall that the Bose-Fermi Kondo model has been analyzed in detail
in the context of quantum critical points of certain heavy fermion materials\cite{Si,Zhu,Demler}, but to our knowledge Eq.\ (\ref{hami}) would 
constitute its first
realization in mesoscopic structures. We get the following Renormalization 
Group (RG) equations akin to those of Zhu and Si\cite{Zhu} and 
Zarand and Demler\cite{Demler} (for $\epsilon=0$)
\begin{eqnarray}
\label{flow}
\frac{d\lambda_{\perp}}{dl} &=& \lambda_{\perp}\lambda_z-\frac{\nu}{2}
\lambda_{\perp}{\gamma}^2\\ \nonumber
\frac{d\lambda_z}{dl} &=& {\lambda_{\perp}}^2\\ \nonumber
\frac{d\gamma}{dl} &=& -\frac{1}{2}\gamma {\lambda_{\perp}}^2, \\ \nonumber
\end{eqnarray}
$l=\ln\left(\Lambda_o/\Lambda\right)$ denotes the scaling variable with
$\Lambda_o\approx \hbox{min}\{E_c,\hbar\omega_c\}$ the initial value of 
the energy cutoff and $\Lambda_o=E_c$ when $R=0$ $(L=0)$.
Our bare values obey $\lambda_{\perp}=J_{\perp}\rho=2|t|\rho$, 
$\lambda_z=J_z\rho=0$, $\gamma^2=R/R_K<1$; $J_z s^z S^z$ refers to the
induced Ising part of the Fermi Kondo coupling\cite{Matveev}, and 
$\nu$ which is of the order of 1 refers to
the normalization constant occurring in the boson time propagator. 
Note that here there is no generated transverse Kondo 
part for the bosonic coupling.  
It is
obvious that since the bosonic heat bath adjusts itself to a given 
(pseudo-)spin configuration this tends to suppress the renormalization of the
Kondo process, i.e., 
the coherent tunneling of electrons from lead to dot. The
competition between the bosonic and fermionic fields then will give rise to 
{\it two}
stable quantum phases corresponding to a {\it Kondo Fermi-liquid phase} 
(where the Kondo energy scale will depend on R) and to
a {\it merely bosonic Ising} regime.

It is
advantageous to rewrite the RG equations above in terms of the two variables
$\lambda_{\perp}$ and
$\hat{\lambda}_z=(\lambda_z-(\nu/2)\gamma^2)$. As long 
as $\gamma^2\ll 1$, this enables us to recover the
renowned Kosterlitz-Thouless equation flow: $d\lambda_{\perp}/dl=\lambda_{\perp}\hat{\lambda}_z$ and $d\hat{\lambda}_z/dl={\lambda_{\perp}}^2$. Hence, this 
strongly validates the idea of a critical  
value for the external resistance, 
namely $R_c[|t|\rho]\approx
4|t|\rho R_K/\nu$, at which the Kondo physics will be 
washed 
out. Let us underline that this formula is only strictly applicable 
in the case of a low-transparency tunnel barrier where $R_c[|t|\rho]\ll R_K$, 
i.e., in the scope of validity of the perturbative RG. 
For $R<4|t|\rho R_K/\nu$, since $\lambda_{\perp}$ and $\hat{\lambda}_z$ will both grow under 
renormalization a Kondo phase with restoration of the SU(2) symmetry 
emerges. The correction to the 
classical capacitance $C=\partial\langle Q\rangle/\partial V_g$ is proportional
to the impurity susceptibility $\chi=\partial\langle S^z\rangle/\partial h$.
An important point is that when applying a strong magnetic field the fixed
point corresponds to the usual one-channel Kondo Fermi liquid where
$\chi=(T_K[R])^{-1}$, and the Kondo energy 
scale can be approximated by the one of the
completely symmetrical model $T_K[R]\approx 
\Lambda_o\exp[-1/(\lambda_{\perp}-\nu\gamma^2/2)]\ll E_c$. 

\begin{figure}[ht]
\centerline{\epsfig{file=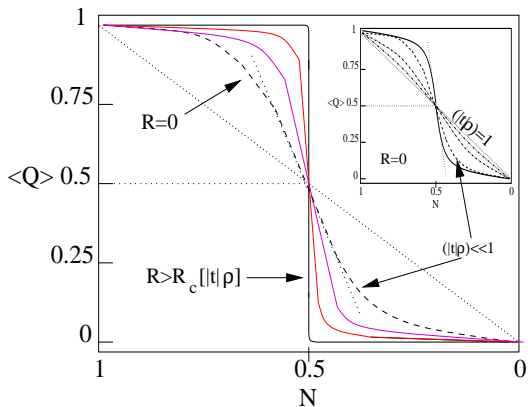,angle=0.0,height=5.3cm,width=6.9cm}}
\vskip -0.2cm
\caption{Average dot's charge $\langle Q\rangle$ 
versus the rescaled voltage $N=V_g C_g/e$ for 
{\it weak} tunneling $(|t|\rho)\ll 1$ and in the case of a single-mode 
spin-polarized point contact. When $R$ will exceed 
$R_c[|t|\rho]\approx 4|t|\rho R_K/\nu$, we predict the
restoration of perfect Coulomb steps\cite{note2}. (inset) Evolution of
the Coulomb blockade oscillations for $R=0$ and various dot-lead couplings.}
\end{figure}

Close to the degeneracy point $N=1/2$, the average dot's charge 
$\langle Q\rangle$ then 
varies linearly with the applied gate voltage $V_g$
at zero temperature and the slope 
$C=\partial\langle Q\rangle/\partial V_g$ will progressively 
increase with R (see Fig. 2).
When $R\geq 4|t|\rho R_K/\nu$, the Kondo temperature vanishes which means 
that $\lambda_{\perp}(l)\rightarrow 0$ and that $\gamma$
remains almost unrenormalized.
Since $\langle \Phi\rangle=0$, we expect a perfect jump in 
$\langle Q\rangle$ or $\langle S^z\rangle$ similar to a free spin in magnetic 
field: $\langle Q\rangle=1$ if $h>0$ $(N>1/2)$ and
$\langle Q\rangle=0$ if $h<0$ $(N<1/2)$. This shows that 
a reasonably large resistance in the electrical circuit
can markedly restore a perfect Coulomb staircase behavior by suppressing
the virtual tunneling of electrons between the lead and the metallic 
dot. Note that an analogous phenomenon has been reported 
in the different context of a {\it small} ``noisy''
dot (embodied by the lowest unoccupied electron level 
as opposed to our many-body grain) 
embedded in a {\it small} mesoscopic
ring (described by the topmost occupied electron level as opposed to our
semi-infinite lead)\cite{Markus}. In the case of 
spinful fermions, i.e., without magnetic field, the RG flow would be 
quite similar 
to that of Eq.~(\ref{flow}). However, the effect of the external 
impedance should be much less spectacular because already in the Kondo phase 
there is a blatant divergence in the capacitance for 
$N\approx 1/2$\cite{Matveev}
\begin{equation}
C\propto -\frac{2E_c}{|t|\rho}\exp\left(\frac{1}{2|t|\rho}\right)
\ln\left(\frac{1}{|2N-1|}\right)\cos(2\pi N),
\end{equation} 
emerging clearly 
from the two-channel Kondo model (the two channels being the spin
polarizations of an electron).

A maybe more physical way of understanding the effect of the electrical 
environment is to absorb $\delta V_g$ in the tunneling term using a unitary 
transformation\cite{note} following 
Refs.~\onlinecite{Nazarov2,Devoret}. In our situation, $H_t$
naturally turns into:
\begin{equation}\label{phase}
\hat{H}_t=J_{\perp}e^{i\varphi}s^+S^-+h.c.,
\end{equation}
where the phase 
$\varphi(t)=(e/\hbar)\int_{-\infty}^{t} \delta V_g(t')dt'$ at the time $t$ 
represents the integral over the voltage fluctuations felt by the 
dot\cite{note0}.  
{\it 
It becomes (more) transparent that a transferred charge onto the dot must be 
accompanied by zero-point fluctuations (excitations) of the environment.}
In this formulation of the tunneling Hamiltonian, the Kondo coupling 
$J_z$ will be renormalized by the contribution
\begin{equation}
\label{bath2}
{J_{\perp}}^2\int d\tau\ sgn(\tau)\ G(\tau)\ e^{K(\tau)},
\end{equation}
$G(\tau)=1/\tau$ corresponds to the usual fermionic imaginary time 
propagator. The main difference with the original Kondo problem is the occurrence of the phase correlator $K(\tau)=\langle \varphi(\tau)\varphi(0)-
\varphi(0)^2\rangle$. An exact calculation of $K(\tau)$ can be performed
at zero temperature for the transmission line, i.e., for the bath of harmonic
oscillators, and for large $\tau=it$ we find  
$K(\tau)=-2(R/R_K)\ln\left(\omega_c|\tau|\right)$\cite{Devoret,Markus}. When $R$ is negligible,
$K(\tau)\approx 0$, one easily recovers the second line of Eq.~(\ref{flow}) 
and then the expected one-channel Kondo fixed point. By increasing $R$, the integral 
$\int_{\hbar/\Lambda_o}^{\hbar/\Lambda} d\tau\ \tau^{-2\alpha-1}$
involved in the above equation (\ref{bath2}) will remain very small 
whatever the 
energy $\Lambda$. Hence, this leads to the same conclusion as previously: $J_z(l)$ ($J_{\perp}(l)$) will persist to be
insignificant up to zero temperature and then
the (pseudo-)spin $S^z$ will satisfy $\langle S^z\rangle=\pm 1/2$. 
This can be entirely attributed to the fact that, for large $R$, 
the probablity $P(\tau)=e^{K(\tau)}$ to observe zero-point fluctuations
in the environment 
rapidly decreases for a long time $\tau$\cite{Nazarov2,Devoret} and 
as a result long-time
(low-energy) Kondo (tunneling) processes 
will be eliminated by the noise.

Now, we shall concentrate 
on the opposite limit, i.e., close to perfect transmission. 
Here, electronic wave packets are clearly
spread out between the bulk lead and the granule, and hence one could 
anticipate
that the zero-point fluctuations of the environment will have more 
difficulty to affect the charge fluctuations of the quantum box.  
To make it more quantitative, we can
resort to bosonization and tackle the one-dimensional problem emerging at 
the QPC along the lines of Refs.~\onlinecite{Matveev,Georg}. Consequently, 
the Coulomb term becomes
\begin{equation}
H_c=\frac{E_c}{\pi}\left(\phi(0)-\sqrt{\pi}N\right)^2-E_cN^2
-\frac{e}{\sqrt{\pi}}\phi(0)\delta V_g,
\end{equation}
where the boson $e\phi(0)/\sqrt{\pi}$ describes the charge on the granule 
and the coordinate $x=0$ refers to the entrance of the dot (or QPC). First, we 
already note that exactly at 
perfect transmission $(\rho|t|=1)$, as long as $\langle \delta V_g\rangle=0$,
the capacitive coupling with the back gate will definitely impose
$\langle \phi(0) \rangle \approx \sqrt{\pi}N$, and then the grain's charge 
will vary linearly
with the (mean) back-gate voltage whatever the external resistance (inset in Fig. 2). Obviously,
the previous qubit basis $S^z=\pm 1/2$ would not be appropriate in that limit.
The weak backscattering contribution at the QPC then can be manipulated 
as\cite{Matveev}
\begin{equation}
H_{bs}=-\frac{v_F}{\pi a}|r|\cos\left(\sqrt{4\pi}\delta\phi(0)\right)
\cos\left(2\pi N\right),
\end{equation}
$a$ is a short-distance cutoff, $v_F$ denotes the Fermi velocity, 
the small reflection amplitude obeys 
$|r|=(1-(\rho|t|)^2)^{1/2}\ll 1$, and the charge 
fluctuation 
field is $\delta\phi(0)=\phi(0)-\sqrt{\pi}N$. For spinless fermions, the
(first-order) correction to the ground state
energy naturally takes the form
$\delta E_1=-v_F/(\pi a)|r|e^{-2\pi\langle \delta\phi(0)^2\rangle}
\cos\left(2\pi N\right)$.

Let us first recall that when $R/R_K\rightarrow 0$, 
$\langle \delta\phi(0)^2\rangle$ is large but finite, and a simple
calculation provides
$\langle \delta\phi(0)^2\rangle=-1/(2\pi)\ln\left(a\mu E_c/v_F\right)$
where $\mu=e^{\cal{C}}$ and $\cal{C}\approx$ 0.5772 is the Euler's 
constant; the phonons with energies below $E_c$ are pinned down by the
interaction term. The periodic correction to the ground-state energy then reads
\begin{equation}
\label{osc}
\delta E_1= -\frac{|r|}{\pi}\mu E_c\cos\left(2\pi N\right).\\ \nonumber
\end{equation}
The average charge in the dot is given by
$e\langle Q\rangle=eN-
e/(2E_c)\partial\delta E_1/\partial N=eN-\mu e|r|\sin\left(2\pi N\right)$. 
One recovers a curve similar to the dashed line in Fig. 2 even though the 
amplitude of the charge oscillations is much smaller 
(i.e., this is much closer to the dotted straight line). 
Now, we revisit the quantum fluctuations of the charge displacement 
for larger external impedances. 
The part of the Coulomb term containing $\delta\phi(0)$
can be turned into
\begin{equation}
H_c=\frac{E_c}{\pi}{\delta\phi(0)}^2-e\delta\phi(0)\sum_j \beta_j X_j,
\end{equation}
with $\beta_j=\gamma\lambda_j/\sqrt{\pi}$. It is suitable to note the
resemblance with a Brownian particle in a harmonic potential. By exploiting 
this analogy\cite{CL,Weis}, we can check that
the following correlation function
$\langle \delta\phi(0,t)\delta\phi(0,0) \rangle \propto - R/\left({E_c}^2 t^2\right)$ almost vanishes for long times and hence that 
the noise is rather irrelevant for $|r|\ll 1$. The 
charge displacement $\langle \delta\phi(0)^2\rangle$ will
be still dominated by the logarithmic phonon contribution 
above cutoff at the time $\hbar/E_c$. Zero-point fluctuations of the 
transmission line cannot fundamentally 
affect the result of Eq.~(\ref{osc}), but for very
large and probably unrealistic  
impedances $R/R_K\rightarrow +\infty$, we could
expect that the logarithmic contribution of 
$\langle \delta\phi(0)^2\rangle$ will be cutoff at 
$\tau_c\leq \hbar/E_c$. In the absence of 
the (Coulomb) term $E_c{\delta\phi(0)}^2/\pi$, for long 
times one would still obtain a logarithmic growth of the 
$\delta\phi(0)$ correlation function\cite{Chamon},
$\langle \delta\phi(0,t)\delta\phi(0,0)\rangle\propto 
-(R_K/R)\ln t$. 

More physically, the scattering of electrons at the QPC produces a phase 
shift in the wavefunctions $\psi_k(x)=\cos\left(k|x|-\delta\right)$ which is 
naturally related
to the average charge entering onto the dot through the Friedel
sum rule $\delta=\pi\langle Q\rangle/e$. As long as $|r|\ll 1$, 
this means that we can approximate $\langle Q\rangle/e\approx N$
whatever the external Ohmic resistance (again $\langle \delta N\rangle=0$).
This ensures the following Friedel oscillation of the electron density
$\rho(x)\approx
E_c/(\hbar v_F)\cos(2k_F|x|-2\delta)$ for $|x|\ll v_F/E_c$ with $2\delta \sim 2\pi N$; the restriction 
of wave 
vectors $k$ to the range $E_c/(\hbar v_F)$ around the Fermi momentum $k_F$ is 
because
only for those wavevectors elastic reflection from the contact takes place. 
We can then infer that the shift in
the ground state energy $-\int dx\ \rho(x)V(x)\delta(x)$,
where the potential $V=|r|\hbar v_F$, (roughly) is still equal to $\delta E_1$.

In closing, we have shown that in the context of a metallic granule
and a low-transparency tunnel barrier, zero-point fluctuations of a 
transmission line can alter the quantum tunneling of electrons {\it 
at low energy}
and then restore a perfect Coulomb staircase behavior. A clear observation of 
such a suppression of the quantum charge 
fluctuations 
would necessitate the application of a strong in-plane magnetic 
field, a reasonable external impedance (a non-negligible fraction
of $R_K$), and again a small transmission coefficient between the metal 
lead and the grain. When approaching the perfect transmission,  
quantum charge fluctuations become so prominent such that the 
electrical environment is rather inefficient. 

We acknowledge M. B\"{u}ttiker and P. Simon for discussions. This work was 
supported in part by NSERC.


\begin{thebibliography}{99}
\bibitem{note0}
$\delta Q=C_g\delta V_g$
can be viewed as the sum over the oscillator normal coordinates $X_j$. 
Its conjugate is the generalized flux $(\hbar/e)\varphi$ with
$\varphi(t)=(e/\hbar)\int_{-\infty}^{t} \delta V_g(t')dt'$
which can be expanded in terms of the momenta $P_j$.
\bibitem{note2}
We obtain the same conclusion away from the degeneracy point(s) by using 
perturbation theory of Ref.~\onlinecite{Matveev} and by absorbing
$\delta V_g$ in the tunneling term similar to Eq.\ (\ref{phase}).
\end{thebibliography}
\end{document}